\documentstyle[aps,eqsecnum,epsf,12pt]{revtex}

\def\la{\langle}
\def\ra{\rangle}

\def\da{^\dagger}
\def\nn{\nonumber}

\def\lh{\lambda_h}
\def\lhh{\bar{\lambda}_h}
\def\lc{\lambda_c}
\def\lhc{\bar{\lambda}_c}
\def\lae{\lambda_e}
\def\lhe{\bar{\lambda}_e}

\def\P{\hat{P}}
\def\H{\hat{H}}
\def\G{\hat{\Gamma}}

\newcommand{\beq}{\begin{equation}}
\newcommand{\eeq}{\end{equation}}
\newcommand{\beqa}{\begin{eqnarray}}
\newcommand{\eeqa}{\end{eqnarray}}
\newcommand{\bc}{\begin{center}}
\newcommand{\ec}{\end{center}}
\newcommand{\bit}{\begin{itemize}}
\newcommand{\eit}{\end{itemize}}

\input psfig.sty

\begin{document}

%%%%%%%%%%%%%%%%%%%%%%%%%%%%%%%%%%%%%%%%%%%%%%%%%%%%%%%
\title{Quantum thermodynamic cooling cycle}

\author{Jos\'e P. Palao\footnote{Also: {\it Departamento de 
F\'{\i}sica Fundamental II, Universidad de La Laguna, La Laguna 38204,
Spain}} and Ronnie Kosloff}

\address{Department of Physical Chemistry and the Fritz Haber Research
Center for Molecular Dynamics, Hebrew University, Jerusalem 91904, 
Israel}

\author{\\ Jeffrey M. Gordon}

\address{Department of Energy and Environmental Physics, Blaustein 
Institute for Desert Research, Ben-Gurion University of the Negev, Sede
Boqer Campus 84990, Israel}

\address{and The Pearlstone Center for Aeronautical Engineering Studies, 
Department of Mechanical Engineering, Ben-Gurion University of the 
Negev, Beersheva 84105, Israel}

\maketitle

%%%%%%%%%%%%%%%%%%%%%%%%%%%%%%%%%%%%%%%%%%%%%%%%%%%%%%%%
\begin{abstract}

\noindent

The quantum-mechanical and thermodynamic properties of a 
3-level molecular 
cooling cycle are derived. An inadequacy of earlier models is rectified in
accounting for the spontaneous emission and absorption associated with the
coupling to the coherent driving field via an environmental reservoir. 
This additional coupling need not be dissipative, and can provide a thermal 
driving force - the quantum analog of classical absorption chillers. 
The dependence of the maximum attainable cooling rate on temperature, 
at ultra-low temperatures, is determined and shown to respect 
the recently-established fundamental bound based on the second 
and third laws of thermodynamics.
\\

\noindent{PACS number(s): 05.70.Ln, 32.80.Pj}

\end{abstract}

%%%%%%%%%%%%%%%%%%%%%%%%%%%%%%%%%%%%%%%%%%%%%%%%%%%%%%%%%

\section{Introduction}

Recently, a fundamental model-independent bound was derived 
for the temperature dependence of the fastest rate
$\dot{\cal Q}_c$ at which any substance can be cooled toward 
absolute zero via energetic exchange with a cold reservoir 
at temperature $T_c$ \cite{kosloff00}:
\beq\label{eq:bound}
\dot{\cal Q}_c\propto T_c^\alpha\,,\;\;\;\;\;\;\;\alpha\geq 1\,.
\eeq
In the same study, a 3-level quantum-mechanical cooling 
cycle was postulated and shown to achieve this bound. The notion of 
analyzing molecular cooling with quantum thermodynamic 
cycles had been introduced earlier \cite{geva96}.
A system (working fluid) rejects heat to a hot bath, 
removes heat from a cold bath, and is driven by a coherent 
driving field: an idealized model for the laser cooling 
of magnetically-confined gases at ultra-low temperatures, 
as well as the laser cooling of solids and dyes
\cite{epstein95,clark98}.

An important ingredient absent from previous models is the 
spontaneous emission and absorption related to the coupling 
between the system and the driving field. Superficially, 
this extra coupling provides a dissipative path and hence 
lowers achievable cooling rate. However, we will show that this 
coupling also introduces the possibility of driving the cooling 
cycle thermally, and broadens the conditions under which cooling 
can be generated. It constitutes the quantum-mechanical analog of 
classical absorption chillers. The quantum dynamics and thermodynamic 
properties of this improved 3-level model will be derived, in 
particular: (a) the identification of thermodynamic with quantum-mechanical 
variables; (b) the conditions required to produce cooling; 
(c) the dependence of cooling rate on the quantum control variables; 
(d) the efficiency of the cooling cycle; and (e) rigorous confirmation 
that the fundamental bound of Eq (\ref{eq:bound}) is respected as 
the absolute zero is approached.

%%%%%%%%%%%%%%%%%%%%%%%%%%%%%%%%%%%%%%%%%%%%%%%%%%%%%%%%%%%%%%%%%%%%%%%

\section{The 3-level model}\label{se:3lr}

\begin{figure}[h]
\vspace{1.2cm}
\hspace{0.16\textwidth}
\psfig{figure=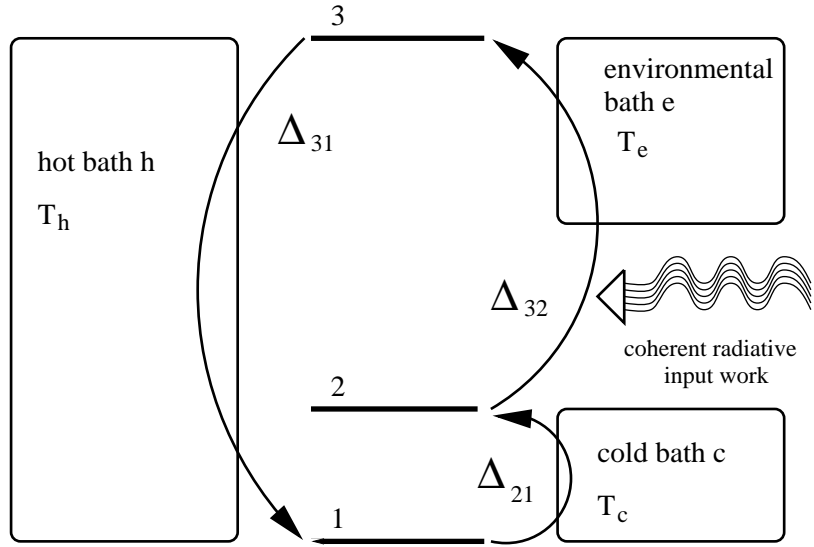,width=0.65\textwidth}
\vspace{.3cm}
\caption{Schematic of the 3-level quantum cooling cycle.}

\label{fig:scheme}
\end{figure}

The model is portrayed schematically in Fig \ref{fig:scheme}:
a 3-level system coupled to 3 infinite baths (reservoirs) 
plus an external driving field of coherent radiation. 
The allowed transitions are: (1) between levels 1 and 3, 
with a hot bath at temperature $T_h$ (heat rejection); 
(2) between levels 1 and 2, with a cold bath at temperature $T_c$ 
(heat removal, i.e., cooling); and (3) between levels 2 and 3, 
simultaneously with the driving field and an environmental bath 
at temperature $T_e$ (dissipative heat rejection when $T_e \le T_h$, 
and heat input when $T_e > T_h$). 
The environmental bath is treated as either: 
(a) independent of the hot and cold baths; or, 
when interference among the transitions is negligible 
\cite{agarwal01}, 
(b) representing an additional transition to the hot or cold bath. 
Two special cases are: (1) $T_e = T_h$, signifying spontaneous emission 
to the hot bath, i.e., coupling the transition to the hot bath; and 
(2) $T_e = T_c$, describing non-radiative (e.g., phonon) decay, i.e., 
coupling to the cold bath.

%%%%%%%%%%%%%%%%%%%%%%%%%%%%%%%%%%%%%%%%%%%%%%%%%%%%%
%%%%%%%%%%%%%%%%%%%%%%%%%%%%%%%%%%%%%%%%%%%%%%%%%%%%%%

The Hamiltonian of the extended system (system, baths and driving field) 
is expressed as
\beq
\H\,=\,\H_s\,+\,\H_{sf}(t)\,+\,\H_{sh}\,+\,\H_{sc}\,+\,\H_{se}\,+
\,\H_{h}\,+\,\H_{c}\,+\,\H_{e}\,
\eeq
where $\H_s$ is the Hamiltonian of the 3-level system 
(the working fluid of the thermodynamic cycle);  
$\H_{sf}(t)$ describes the time-dependent coupling between the system 
and the driving field (coherent radiation);
$\H_{sh}$, $\H_{sc}$, $\H_{se}$ 
account for the coupling between the system and its respective baths; 
and $\H_{h}$, $\H_{c}$, $\H_{e}$ 
are the respective Hamiltonians of the hot ($h$), cold ($c$) 
and environmental ($e$) baths.

The system's Hamiltonian can be cast as
\beq
\H_s\,=\,\Delta_{31}\,\P_{33}\,+\,\Delta_{21}\,\P_{22}\,
\eeq
where $\Delta_{ij}=E_i-E_j$ 
is the transition energy between levels $i$ and $j$ 
(with $E_1$ chosen as zero),
and $\P_{ij}=|i\ra\la j|$
for $i=j$ are the 
projection operators over the states $i=1,2,3$. 
(Our units are chosen such that $\hbar=1$ and $k_B=1$.)

Our expression for $\H_{sf}(t)$ is based on the semi-classical 
rotating wave approximation \cite{cohen92}:
\beq
\H_{sf}(t)\,=\,\epsilon\,(\,\P_{32}\,e^{-i\omega t}\,+
\,\P_{23}\,e^{i\omega t}\,)\,
\eeq
for a field frequency $\omega$ and a coupling strength 
$\epsilon$ which depends on the amplitude of the driving 
field as well as the dipole moment of the transition, 
the latter being assumed to be independent of $\Delta_{32}$. 
With linear coupling between each transition in the 3-level system 
and its corresponding bath, the other coupling Hamiltonians 
can be written as:
\beqa\label{eq:coup}
\H_{sh}&=&\P_{31}\G_h\,+\,\P_{13}\G_h\da\,,\nn\\
\H_{sc}&=&\P_{21}\G_c\,+\,\P_{12}\G_c\da\,,\nn\\
\H_{se}&=&\P_{32}\G_e\,+\,\P_{23}\G_e\da\,,
\eeqa
where $\P_{ij}$ now represents the raising ($i>j$) or lowering ($i<j$) 
operator for the transition $i\leftrightarrow j$, 
and $\G_b$ denotes a bath operator ($b=h,c,e$). 
Only the general properties of the bath correlation 
functions are needed to obtain the reduced description 
of the system's dynamics. The more detailed information on 
the bath Hamiltonians and the operators $\G_b$ will be derived 
afterwards when we examine the explicit forms of the correlation functions.

The standard tools for quantum open systems can be used to 
obtain a reduced description of system dynamics, i.e., 
the time evolution for the system's density matrix 
$\hat{\rho}_s(t)$, in terms of the operators of the 3-level system.
$\hat{\rho}_s(t)={\rm Tr}_B\{\hat{\rho}(t)\}$  
where $\hat{\rho}$ is the density matrix of the extended system and 
${\rm Tr}_B$ denotes the trace over all bath variables.
We invoke the following approximations:
\begin{itemize}

\item Weak coupling between the 3-level system and each bath.

\item Factorization of the initial density matrix of the extended system,
\beq\label{eq:initden}
\hat{\rho}(t=0)=\hat{\rho}_s(t=0)
\otimes\,\hat{\rho}_h\,\otimes\,\hat{\rho}_c\,\otimes\,\hat{\rho}_e\,
\eeq
with the initial density matrix of each bath assumed 
to be in thermal equilibrium,
\beq
\hat{\rho}_b=\frac{e^{-\H_b/T_b}}{{\rm Tr}_b\{\hat{\rho}_b\}}\,
\eeq

\item The Markovian assumption that the bath correlations
\beq\label{eq:corr}
C_{\G_b \G'_b}
={\rm Tr}_b\{\G_b\,\G'_{b}(-t)\,\hat{\rho}_b\}
={\rm Tr}_b\{\G_b\, e^{-i\H_bt}\, 
\G'_b\,e^{i\H_bt}\,\hat{\rho}_b\}\,
\eeq
decay rapidly relative to the time scale of evolution 
of the 3-level system.

\item Weak coupling with the driving field, i.e., $\epsilon$
being of the same order as the coupling with the baths. 
This approximation permits us to treat the effect of the field 
on the baths and the system separately \cite{geva96,cohen92}.

\end{itemize}

These approximations result in the density matrix of the 
extended system factorizing at all times. One consequence 
is that the interfaces between the system and its baths become 
isothermal partitions which allow energy transfer but do not destroy 
the system's integrity, i.e., no quantum entanglement is created between 
the system and its baths \cite{lindblad96}.
This point is essential to the validity of the thermodynamic cycle approach, 
i.e., to distinct energy flows between a system and its reservoirs 
such that the intactness of the system is not compromised.

The analyses that follow build upon the derivation of 
the system's reduced dynamics as detailed in Appendix \ref{ap:rd}.
We adopt the Heisenberg representation and express the time 
evolution of an arbitrary operator $\hat{X}$ as
\beq\label{eq:dhp}
\dot{\hat{X}}\,=\, i [\H_{s}+\H_{sf}(t),\hat{X}]
+(\frac{\partial \hat{X}}{\partial t})\,+
\sum_{b=h,c,e} {\cal L}_b(\hat{X})\,
\eeq
where the super-operator ${\cal L}_b$ describes the effect of the 
baths on the dynamics of the 3-level system and possesses 
a structure (the Lindblad form) that assures the (complete) positivity 
of the reduced dynamics of Eq (\ref{eq:dhp}):
\beq\label{eq:diss}
{\cal L}_b(\hat{X})= \frac{\lambda_b}{2}\{[\P_b,\hat{X}]\P_b\da 
- \P_b[\P_b\da,\hat{X}]\}
+\frac{\bar{\lambda}_b}{2}\{[\P_b\da,\hat{X}]\P_b - 
\P_b\da[\P_b,\hat{X}]\}\,
\eeq
with $\P_b\,=\,\P_{31}$, $\P_{21}$, $\P_{32}$ for $b=h,c,e$, repectively.
The correlation coefficients $\lambda_b$ and $\bar{\lambda}_b$,
defined in Appendix A, depend on the transition energies, 
the bath temperatures and the coupling strength. They appear as 
coefficients in the equations for 
$\P_{33}$, $\P_{22}$, $\P_{11}$, $\P_{32}$ and $\P_{23}$
that follow from Eq (\ref{eq:dhp}) and de-couple from the others:
\beqa\label{eq:rate}
\dot{\P}_{33}&=&-i\,\epsilon\, e^{-i\omega t}\P_{32} +i\,\epsilon\,
e^{i\omega t}\P_{23}-(\lambda_h+\lambda_e)\P_{33}
+\bar{\lambda}_e\P_{22}
+\bar{\lambda}_h\P_{11}\,\nn\\
\dot{\P}_{22}&=&i\,\epsilon\, e^{-i\omega t}\P_{32} -i\,\epsilon\,
e^{i\omega t}\P_{23}+\lambda_e\P_{33}
-(\lambda_c+\bar{\lambda}_e) \P_{22}
+\bar{\lambda}_c \P_{11}\,\nn\\
\dot{\P}_{11}&=&\lambda_h \P_{33}
+\lambda_c P_{22}
+-(\bar{\lambda}_h+\bar{\lambda}_c) \P_{11}\,\nn\\
\dot{\P}_{32}&=&i \Delta_{32}\P_{32}
-i\,\epsilon\, e^{i\omega t}\P_{33} +i\,\epsilon\,e^{i\omega t}\P_{22}
-\frac{1}{2}\{\lambda_h+\lambda_c+\lambda_e+\bar{\lambda}_e\}P_{32}\,
\nn\\
\P_{23}&=&\P_{32}^\dagger\,.
\eeqa

The terms $\P_{ij}$ are related to the diagonal elements of 
the reduced density matrix when $i=j$ 
(${\rm Tr}_s\{\P_{ii}\hat{\rho}_{s}\}=\rho_{ii}$),
and to the off-diagonal elements when $i\ne j$
(${\rm Tr}_s\{P_{ij}\hat{\rho}_{s}\}=\rho_{ji}$). 
Hence Eq (\ref{eq:rate}) represents 
rate equations for the density matrix elements. 
The diagonal and off-diagonal elements are coupled by the driving field. 
The bath is responsible for the self-couplings of both diagonal and 
off-diagonal elements.
$\lambda_b$ and $\bar{\lambda}_b$ 
are then transition probabilities, per unit time, between energy levels. 
For example, $\lambda_h$ is the transition probability of a decay from level 
3 to 1 in which energy is rejected to the hot bath, and $\bar{\lambda}_h$ 
is the probability for an excitation from level 1 to 3 in which energy is 
absorbed from the hot bath.

%%%%%%%%%%%%%%%%%%%%%%%%%%%%%%%%%%%%%%%%%%%%%%%%%%%%%%%%%%%%%%%%%%%%

\section{Identification of the thermodynamic variables}
\label{se:the}

The identity of the thermodynamic energy flows 
in the cooling cycle follows from energy conservation and averaging
\cite{alicki79,geva95}. 
The Hamiltonian $\H_T=\H_{s}+\H_{sf}(t)$ 
is introduced into the evolution equation (\ref{eq:dhp}), 
is multiplied by the initial density matrix, 
and is traced over the variables of the 3-level
system, to yield
\beq\label{eq:exva}
\la\dot{\H_T} \ra\,=\,\la\frac{\partial\H_{sf}(t)}{\partial t}\ra +
\la{\cal L}_h(\H_T)\ra+\la{\cal L}_c(\H_T)\ra
+ \la{\cal L}_e(\H_T)\ra\,.
\eeq
The energy flow associated with the driving field 
(the first term on the RHS of Eq (\ref{eq:exva})) is the power input 
(cycle-averaged work) to the cycle, $\dot{\cal W}$.
The remaining 3 terms are the respective heat flows 
between the system and its baths, $\dot{\cal Q}_h$,
$\dot{\cal Q}_c$, $\dot{\cal Q}_e$.
At steady state, $\la \H_T\ra$ is constant and 
independent of the system's initial state, so
$\la \dot{\H_T}\ra=0$ and Eq (\ref{eq:exva})
can be expressed as the First Law of thermodynamics 
for the thermodynamic cycle:
\beq
\dot{\cal W}+\dot{{\cal Q}}_h+\dot{{\cal Q}}_c+\dot{{\cal Q}}_e=0\,
\eeq
with energy flows into the system defined as positive.

Now we can derive the relation between the cycle's 
thermodynamic variables and the quantum-mechanical parameters. 
At steady state (denoted by the superscript $ss$), 
the general $P$ operators become
\beq\label{eq:dss}
\P_{ij}\rightarrow \P_{ij}^{ss}\,\,e^{i \alpha_{ij} t}\,
\eeq
where $\dot{\P}_{ij}^{ss}=0$, 
$\alpha_{ii}=0$, $\alpha_{32}=\omega$ and the remaining values
of $\P_{ij}^{ss}$ 
follow from introducing (\ref{eq:dss}) into Eq (\ref{eq:rate}) 
and solving the resultant set of linear time-independent equations.
The solutions for $\P_{ij}^{ss}$ turn out to 
be proportional to the identity operator $\hat{1}$,
$\P_{ij}^{ss}=p_{ij}\hat{1}$, where $p_{ij}$ is real when $i=j$ and
complex when $i\ne j$.
From the relation between the $\P$ operators 
and the elements of the density matrix, the steady-state 
expectation values of the populations (i.e., the diagonal elements) 
are constants, $\la \P_{ii}\ra =\la \P_{ii}^{ss}\ra =p_{ii}$, 
and the expectation values of the off-diagonal elements 
$\P_{32}$ and $\P_{23}$ oscillate in time,
$\la \P_{ij}\ra =\la \P_{ij}^{ss}\,\,e^{i \alpha_{ij} t}\ra =
p_{ij}\,\,e^{i \alpha_{ij} t}$.  
Explicit formulae for $p_{ij}$ are provided in Appendix \ref{ap:coe}
for the condition of resonance that is assumed in the 
analysis that follows
\beq
\omega\equiv \Delta_{32}\,.
\eeq

The cycle-average thermodynamic variables of the quantum 
refrigerator can now be expressed as:
\beqa\label{eq:phf}
\dot{\cal W}&=&2\,\epsilon\,\Delta_{32}\,{\rm Imag}[p_{32}]
= \Delta_{32}\epsilon^2\{A(n_c-n_h)+B\} \,\nn\\
\dot{\cal Q}_h&=&\Delta_{31}\, (-\lambda_h\,p_{33}
+\bar{\lambda}_h\,p_{11})
=-\Delta_{31}\{\epsilon^2 A (n_c - n_h)+C(n_c n_e - n_h)\}\,\nn\\  
\dot{\cal Q}_c&=&\Delta_{21}\, (-\lambda_c\,p_{22}
+\bar{\lambda}_c\,p_{11})
=\Delta_{21}\{\epsilon^2 A (n_c - n_h)+C(n_c n_e - n_h)\}\,\nn\\
\dot{\cal Q}_e&=&\Delta_{32}\, (-\lambda_e\,p_{33}
+\bar{\lambda}_e\,p_{22})
=-\Delta_{32}\{\epsilon^2 B - C(n_c n_e - n_h)\}\,
\eeqa
where Eq (\ref{eq:rel1}) of Appendix \ref{ap:rd} has been used,
and $n_b$ denotes the equilibrium populations:
$n_h=e^{-\Delta_{31}/T_h}$, $n_c=e^{-\Delta_{21}/T_c}$,
$n_e=e^{-\Delta_{32}/T_e}$. 
Detailed expressions for the positive coefficients $A$, $B$ and $C$ 
are presented in Appendix B, and depend on the bath temperatures, 
the transition energies and the coupling strength.
Each energy flow constitutes a competition between two processes with rates
$\Delta_b\, \lambda_b\, p_{kk}$ and 
$\Delta_b\, \bar{\lambda}_b\, p_{ll}$ ($k>l$),
with the former describing heat rejection of magnitude 
$\Delta_b$ to the bath in the decay $k \rightarrow l$, 
and the latter representing heat removal of the 
same magnitude from the bath in the excitation $l \rightarrow k$.
Also, as will be elucidated in Section \ref{se:qac}, 
the work input rate $\dot{\cal W}$ need not be the only driving 
force for the cooling cycle. When $T_e > T_h$, the 
incoherent thermal flow $\dot{\cal Q}_e$ can become positive 
and hence can also contribute to the cooling rate.

The entropy production rate for the cycle is
\beqa\label{eq:ent}
\dot{S}&=&-\{ \frac{\dot{\cal Q}_h}{T_h}+  
\frac{\dot{\cal Q}_c}{T_c} + \frac{\dot{\cal Q}_e}{T_e} \} \,=\nn\\
&=&-4\epsilon^2\frac{\lc \lh}{D} (n_c-n_h)
(\frac{\Delta_{21}}{T_c}-\frac{\Delta_{31}}{T_h})
+4\epsilon^2\frac{\lae (\lhc+\lhh)}{D}\,\frac{\Delta_{32}}{T_e}(1-n_e)+\nn\\
&-&\frac{\lh \lc \lae}{D} (n_h-n_c n_e)
(\frac{\Delta_{31}}{T_h}-\frac{\Delta_{21}}{T_c}-
\frac{\Delta_{32}}{T_e})\,
\eeqa
where $D$ is a positive function derived in Appendix \ref{ap:coe}. 
With the relations noted above for the equilibrium 
populations $n_b$, it is straightforward to prove that $\dot{S}$  
in Eq (\ref{eq:ent}) must be non-negative, in accordance with the Second Law.

%%%%%%%%%%%%%%%%%%%%%%%%%%%%%%%%%%%%%%%%%%%%%%%%%%%%%%%%%%%%%%%%%%

\section{Coupling with the baths}\label{se:bm}

The results for the cooling performance of 
the quantum refrigeration cycle depend on the particulars 
of the coupling between the system and its baths. 
Hence specific models must be invoked. The primary variable of 
interest is the cooling rate (the interaction with the cold bath). 
Since in any event the results that will now be derived are independent 
of the nature of the hot and environmental reservoirs, we will 
treat these two as white baths, i.e., baths with a constant density 
of energy modes (at least in the frequency range of interest), for which

\beq\label{eq:whiteb}
\lambda_b= \Lambda_b\,,\;\;\;\;\;\;
\bar{\lambda}_b=\Lambda_b\,e^{-\Delta_b/T_b}\,
\eeq
where $\Lambda_b$ is the strength of the coupling. 

Earlier quantum refrigeration models also treated the 
cold bath as having a constant mode density 
\cite{kosloff00,geva96}.
Here a more elaborate and realistic model is introduced 
for the cold bath: an assembly of harmonic oscillators 
- a viable model in the weak coupling limit \cite{prokof00}. 
The bath's Hamiltonian can then be expressed as
\beq
\H_c=\sum_i \Delta_{ci}\hat{a}_{ci}\da \hat{a}_{ci}
\eeq
where $\hat{a}_{ci}$ and $\hat{a}_{ci}\da$ 
are the annihilation and creation operators, 
with the index $i$ spanning the bath oscillator energies $\Delta_{ci}$.
The operator of Eq (\ref{eq:coup}), that couples the system and 
bath is then
\beq
\G_c=\sum_i g_{ci}\hat{a}_{ci} + g_{ci}^* \hat{a}_{ci}\da\,
\eeq
where $g_{ci}$ and $g_{ci}^*$ denote the coupling constants.

The influence of an harmonic bath on system dynamics is embedded in the spectral 
strength function \cite{prokof00,vankampen97} 
$J_c(\Delta)=\sum_i |g_{ci}|^2 \delta(\Delta-\Delta_{ci})$
which, for low energies, is well approximated by a simple power law dependence
\beq\label{eq:sfunc}
J_c(\Delta)\approx \Lambda_c \,\,\Delta^{s_c}\,
\eeq
where $s_c$ must be positive \cite{legget87}. The coefficients $\lambda$ 
then follow as
\beq\label{eq:lambda}
\lambda_c(\Delta_c,T_c)=
\frac{\Lambda_c \,\Delta_c^{s_c}}{1-e^{-\Delta_c/T_c}}\,,
\;\;\;\;\;\;
\bar{\lambda}_c(\Delta_c,T_c)=
\frac{\Lambda_c \,\Delta_c^{s_c}}{e^{\Delta_c/T_c}-1}\,.
\eeq
%

%%%%%%%%%%%%%%%%%%%%%%%%%%%%%%%%%%%%%%%%%%%%%%%%%%%%%%%%%%%%%%%%%%

\section{Cooling performance}\label{se:cr}

\subsection{Objectives}

Several aspects of the thermodynamic performance 
of the quantum cooling cycle will be addressed in 
this section: (1) how strong a coupling with the driving 
field is required to achieve cooling 
(i.e., to insure $\dot{\cal Q}_c > 0$ ); 
(2) how does cooling rate vary with the natural control 
variable $\Delta_{21}$ and under what conditions is cooling rate maximized; 
(3) how does maximum cooling rate depend on $T_c$ in the limit of 
the absolute zero and how does this dependence compare to 
the fundamental bound established from the Second and Third Laws; 
and (4) what is the efficiency of this cooling cycle and can it 
be cast in the same form as classical chiller analyses. 
In this section, only $T_e$ values in the range 
$T_c \le T_e \le T_h$ are analyzed, 
so the environmental bath serves a solely dissipative role. 
The following section will focus on the special thermodynamic 
consequences when $T_e > T_h$.

\subsection{Cooling window and electronic analog}

Because of the competition between the coupling to 
the driving field and dissipative losses, only certain 
combinations of system parameters produce cooling. The 
thermoelectric chiller offers a familiar example of the 
existence of a refrigeration window, where the cooling 
effect exists only for: (a) a sufficiently high thermopower 
(Seebeck coefficient); and (b) a particular voltage window. 
There are analogous limits in the 3-level quantum chiller.

In fact, the 3-level model can be viewed as an electronic 
device, with the analog of the voltage $V$ being $\Delta_{32}$ 
(in units of electronic charge). The maximum voltage is 
$V_{\rm max}=\Delta_{31}$, and $\Delta_{21}=V_{\rm max}-V$. 
The power input is the product of voltage and current, 
so that Eq (\ref{eq:phf}) provides the analog of the electrical 
current as a complicated implicit function of the system parameters, 
as well the relation between current and voltage. In the absence 
of the environmental bath, current is an exponentially increasing 
function of voltage \cite{kosloff00} which, not coincidentally, is the same as 
for ideal diodes. With the added dissipation to the environmental 
bath, the current remains a strongly increasing function of voltage, 
albeit not strictly exponential, similar to non-ideal diodes.

\begin{figure}[h]
\vspace{1.2cm}
\hspace{0.16\textwidth}
\psfig{figure=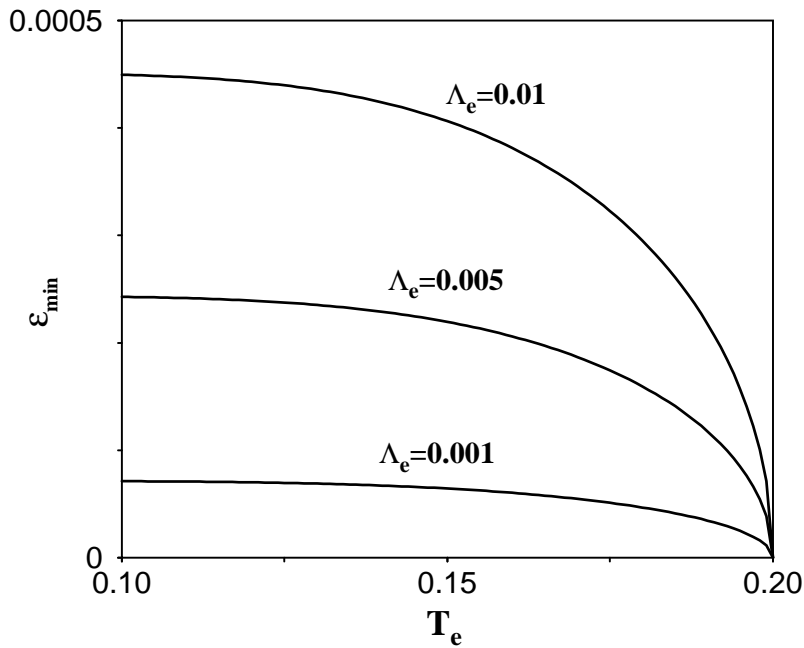,width=0.65\textwidth}
\vspace{.3cm}
\caption{Minimum coupling needed for cooling, 
as a function of the environmental bath temperature, 
at 3 values of coupling coefficient with the environmental bath. 
Other system parameters are fixed at: 
$T_c=0.1$, $T_h=0.2$, $\Delta_{31}=1$, 
$\Lambda_h=\Lambda_c=0.001$, and $s_c=1$.
The regime of $T_e>T_h$, where no coupling is required 
to produce cooling, is addressed in Section \ref{se:qac}.
}
\label{fig:mep}
\end{figure}

The minimum coupling strength with the driving 
field $\epsilon_{\rm min}$ to produce a cooling effect follows 
from Eq (\ref{eq:phf}):
\beq\label{eq:con1}
\epsilon_{\rm min}=\sqrt{C (n_h - n_e)/\{A (1-n_h)\}}\,
\eeq
and is plotted as a function of $T_e$ in Fig \ref{fig:mep}.

%%%%%%%%%%%%%%%%%%%%%%%%%%%%%%%%%%%%%%%%%%%%%%%%%%%%%
%%%%%%%%%%%%%%%%%%%%%%%%%%%%%%%%%%%%%%%%%%%%%%%%%%%%%%

\subsection{Maximum cooling rate and the fundamental bound}

\begin{figure}[h]
\vspace{1.2cm}
\hspace{0.16\textwidth}
\psfig{figure=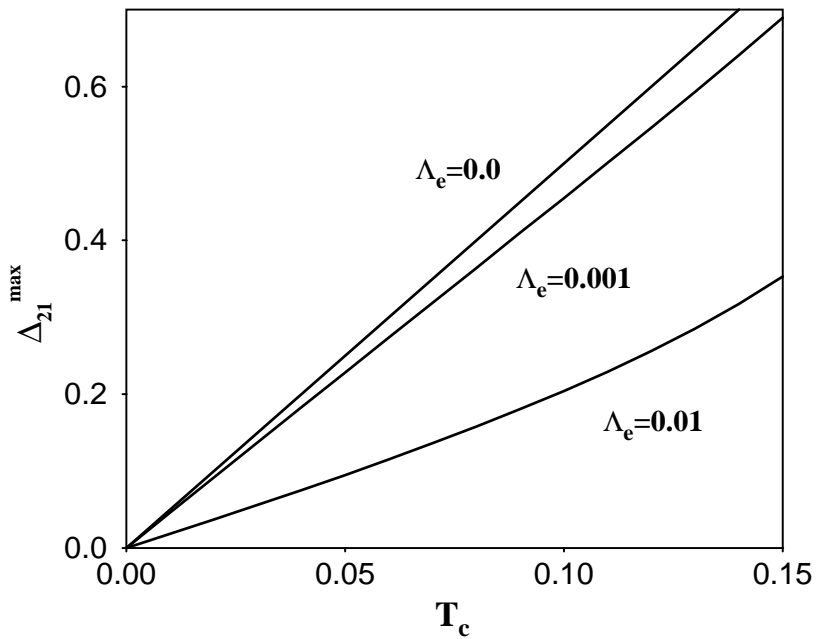,width=0.65\textwidth}
\vspace{.3cm}
\caption{Refrigeration window as a function of $T_c$ 
for 3 values of the coupling coefficient with the 
environmental bath. Other system parameters are fixed at: 
$T_h=T_e=0.2$, $\Delta_{31}=1$,  
$\Lambda_h=\Lambda_c=\epsilon=0.001$ and $s_c=1$.}
\label{fig:rwi}
\end{figure}

\begin{figure}[h]
\vspace{1.2cm}
\hspace{0.16\textwidth}
\psfig{figure=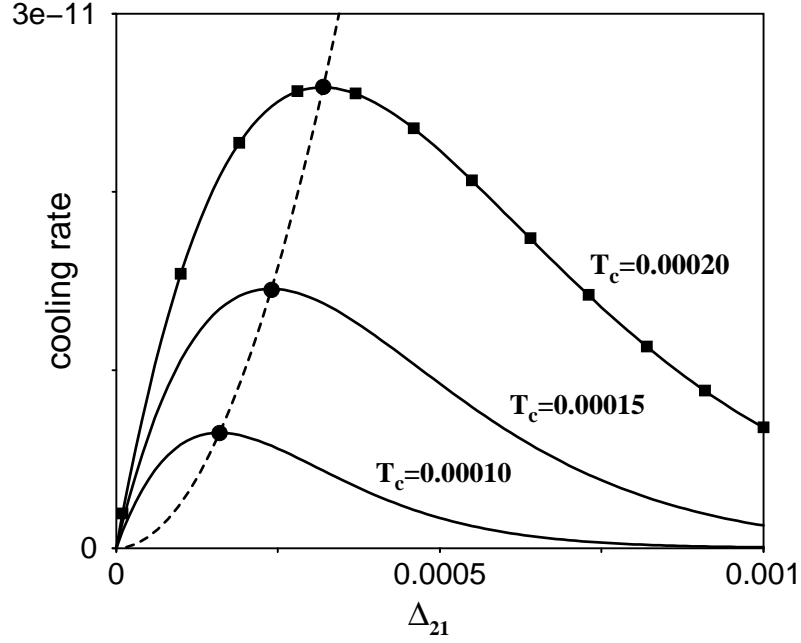,width=0.65\textwidth}
\vspace{.3cm}
\caption{Cooling rate as a function of $\Delta_{21}$ for 3 values 
of cold bath temperature. Other system parameters are fixed at:
$T_h=T_e=0.03$, $\Delta_{31}=1$,
$\Lambda_h=\Lambda_c=\Lambda_e=\epsilon=0.001$ and
$s_c=1$.
The broken curve and solid circles indicate 
the locus of maximum cooling rate. The solid curves are 
nominally exact numerical results, while the solid squares 
illustrate the accuracy of the approximation of Eq (\ref{eq:aprox}).
}
\label{fig:crate}
\end{figure}

Our ``control knob'' for varying cooling rate is the 
transition energy $\Delta_{21}$. The cooling rate vanishes at 
two values of $\Delta_{21}$ that delimit the cooling window: 
$\Delta_{21}=0$ and $\Delta_{21}=\Delta_{21}^{\rm max}$, 
where $\Delta_{21}^{\rm max}$ depends on the principal system 
parameters, most notably $T_c$ and $\Lambda_e$ (Eq (\ref{eq:phf})), 
as illustrated in Fig \ref{fig:rwi}. Sample curves of cooling rate as 
a function of $\Delta_{21}$ are plotted in Fig \ref{fig:crate}.

%%%%%%%%%%%%%%%%%%%%%%%%%%%%%%%%%%%%%%%%%%%%%%%%%%%%%
%%%%%%%%%%%%%%%%%%%%%%%%%%%%%%%%%%%%%%%%%%%%%%%%%%%%%%

%%%%%%%%%%%%%%%%%%%%%%%%%%%%%%%%%%%%%%%%%%%%%%%%%%%%%
%%%%%%%%%%%%%%%%%%%%%%%%%%%%%%%%%%%%%%%%%%%%%%%%%%%%%%

Solving for the maximum cooling rate numerically 
in the limit $T_c\rightarrow 0$ 
reveals that this 3-level quantum 
chiller respects the fundamental bound of Eq (\ref{eq:bound}). 
A more satisfying analytic derivation is possible once 
we establish the accuracy of an approximate closed-form 
expression for the cooling rate in the limit of the absolute zero. 
In the limit of vanishing $T_c$: (a) $\Delta_{21}^{\rm max}$ vanishes; 
(b) $\Delta_{32}\approx\Delta_{31}$; and (c) 
the coefficients $c_{10}$, $c_{20}$ and $c_{d0}$ 
of Appendix \ref{ap:coe} become independent of both $T_c$ 
and $\Delta_{21}$. Then, to an excellent approximation, the 
cooling rate reduces to
\beq\label{eq:aprox}
\dot{\cal Q}_c\approx \Delta_{21}^{(s_c+1)}\,\,\,\,
\frac{\Lambda_c}{1-e^{-\Delta_{21}/T_c}}\,\,\,\, 
\frac{(e^{-\Delta_{21}/T_c}\,c_{10}-c_{20})}
{c_{d0}}\,.
\eeq
The accuracy of this approximation is depicted in Fig \ref{fig:crate}. 
Although maximizing the cooling rate of Eq (\ref{eq:aprox}) yields 
a transcendental equation that can only be solved numerically, 
the solution for that transition energy $\Delta_{21}^*$ is of the 
form $\Delta_{21}^*\propto T_c$, independent of the coupling with 
the environmental bath or the value of the positive exponent $s_c$ 
in Eq (\ref{eq:sfunc}). It then follows from Eq (\ref{eq:aprox}) that
\beq\label{eq:mcr}
\dot{\cal Q}_c^{\rm max}\propto T_c^{(s_c+1)}\;\;\;\;\;\;\;\;(s_c>0)\,
\eeq
as required by the Second and Third laws \cite{kosloff00}.

\subsection{Efficiency}

%%%%%%%%%%%%%%%%%%%%%%%%%%%%%%%%%%%%%%%%%%%%%%%%%%%%%
\begin{figure}[h]
\vspace{1.2cm}
\hspace{0.16\textwidth}
\psfig{figure=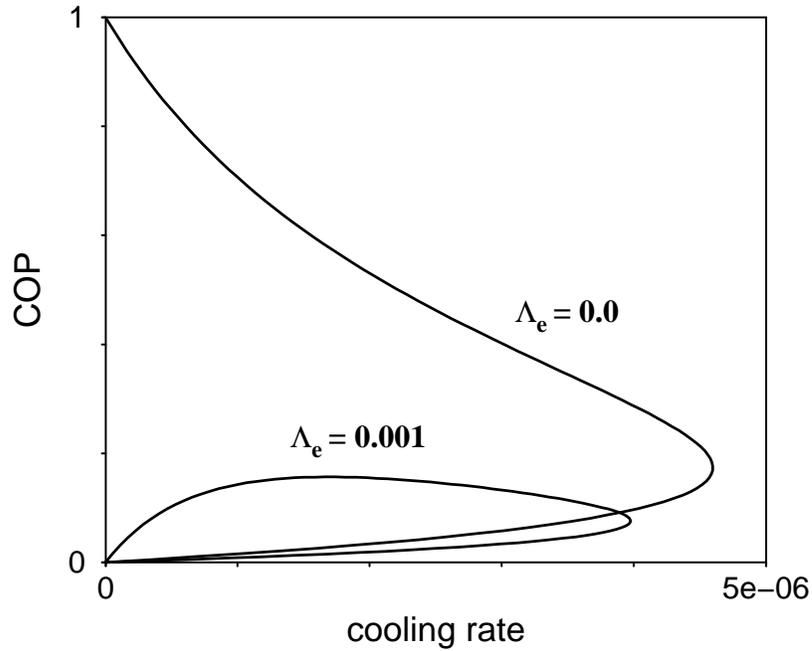,width=0.65\textwidth}
\vspace{.3cm}
\caption{Characteristic chiller plot, {\rm COP} against cooling rate, 
with and without coupling to the environmental bath. 
The control variable is $\Delta_{21}$. System parameters 
are fixed at: $T_c=0.1$, $T_h=T_e=0.2$,
$\Delta_{31}=1$,  
$\Lambda_h=\Lambda_c=\epsilon=0.001$ and $s_c=1$.
}
\label{fig:cop1}
\end{figure}
%%%%%%%%%%%%%%%%%%%%%%%%%%%%%%%%%%%%%%%%%%%%%%%%%%%%%%

\begin{figure}[h]
\vspace{1.2cm}
\hspace{0.16\textwidth}
\psfig{figure=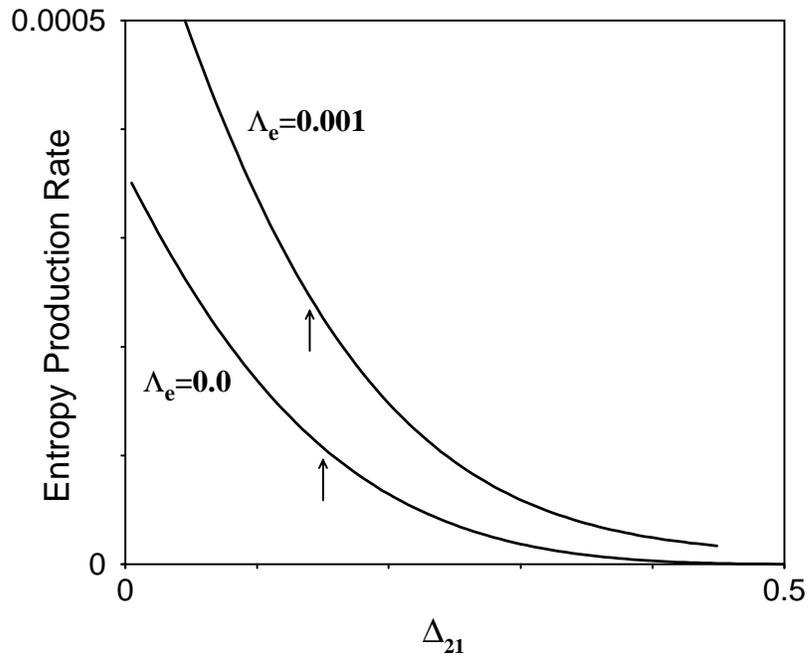,width=0.65\textwidth}
\vspace{.3cm}
\caption{Entropy production rate as a function of 
$\Delta_{21}$, with and without coupling to the environmental bath. 
The refrigeration window is broader in the absence of 
this coupling (the two curves terminate at different 
values of $\Delta_{21}^{\rm max}$). 
Arrows indicate the points of maximum cooling rate. 
System parameters are fixed at: $T_c=0.1$,
$T_h=T_e=0.2$, $\Delta_{31}=1$,    
$\Lambda_h=\Lambda_c=\epsilon=0.001$ and $s_c=1$.
}
\label{fig:entp}
\end{figure}

Cooling cycle efficiency is usually defined 
by Coefficient of Performance (${\rm COP}$), the ratio of 
cooling rate to input power, which in this instance is 
${\rm COP}=\dot{\cal Q}_c/\dot{\cal W}$. 
Cooling cycles are conveniently characterized by a plot of 
COP against cooling rate \cite{gordon00}, 
as in Fig \ref{fig:cop1}. Even in the absence 
of the (parasitic) environmental bath, the 3-level system 
possesses an energy leak that militates against efficient 
operation as  $\Delta_{21}\rightarrow 0$ \cite{kosloff00}, 
which appears as the lower branch of the curves in 
Fig \ref{fig:cop1}, with ${\rm COP}$ vanishing as cooling 
rate is lowered. At the other end of the refrigeration window
($\Delta_{21}\rightarrow \Delta_{21}^{\rm max}$), 
the existence of the extra dissipation to the environmental 
bath makes a quantum difference. 
In the absence of this extra energy-leak pathway, 
there are no irreversibilities that mitigate against efficiency 
operation, so both cooling rate and power input vanish at the same 
rate such that the ${\rm COP}$ approaches its fundamental 
reversible (Carnot) value of $T_c/(T_h - T_c)$ 
\cite{kosloff00,gordon00}. The coupling to the environmental 
bath introduces a loss mechanism that mitigates against this 
nominally slow operation, so {\rm COP} also vanishes in this limit 
(the upper branch of the curve in Fig \ref{fig:cop1} for 
$\Lambda_e > 0$). Fig \ref{fig:entp} offers an alternative 
view of the differences in dissipation: a plot of entropy 
production rate against $\Delta_{21}$ with and without 
coupling to the environmental bath.

%%%%%%%%%%%%%%%%%%%%%%%%%%%%%%%%%%%%%%%%%%%%%%%%%%%%%
%%%%%%%%%%%%%%%%%%%%%%%%%%%%%%%%%%%%%%%%%%%%%%%%%%%%%%

%%%%%%%%%%%%%%%%%%%%%%%%%%%%%%%%%%%%%%%%%%%%%%%%%%%%%%%%%%%%%%%%%%%

\section{Quantum absorption chiller}\label{se:qac}

Classical cooling cycles can be driven by thermal 
sources that are hotter than the heat rejection reservoir, 
commonly called absorption chillers (in contrast to more 
common work-driven mechanical chillers) \cite{gordon00}. 
A simple quantum analog is our 3-level cooling cycle with 
$T_e > T_h$, where the 
environmental bath is analogous to what is called the generator 
in classical absorption cycles. Consider de-coupling the 3-level 
system from the coherent radiation field and driving it with 
incoherent radiation (e.g., a flash lamp) at input thermal power 
$\dot{\cal Q}_e$. The environmental bath becomes a heat source 
rather than a dissipative sink. As in classical absorption chillers, 
the {\rm COP} is defined as 
$\dot{\cal Q}_c/\dot{\cal Q}_e$. 
While the refrigeration window and 
characteristic chiller curve will now be derived, we note 
that the basic result for the dependence of maximum cooling 
rate on $T_c$ in the limit $T_c\rightarrow 0$ 
(\ref{eq:mcr}) remains unaltered.

The refrigeration window follows from Eq (\ref{eq:phf}):
\beq\label{eq:dmhp}
\Delta_{21}^{\rm max}=\frac{T_c\,(T_e-T_h)}{T_h\,(T_e-T_c)}\,\Delta_{31}\,
\eeq
which, as for the dissipative environmental bath 
at $T_e\le T_h$, vanishes in the limit $T_c \rightarrow 0$.

\begin{figure}[h]
\vspace{1.2cm}
\hspace{0.16\textwidth}
\psfig{figure=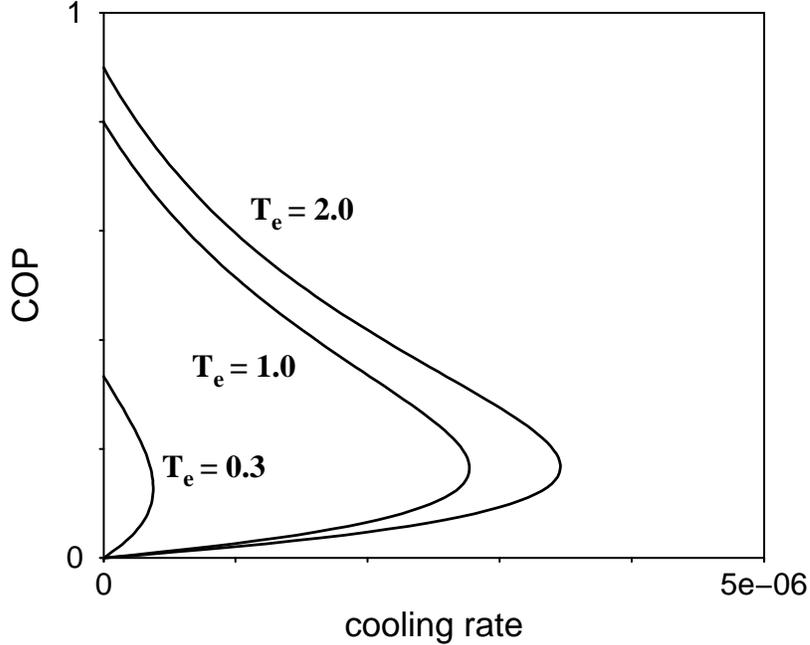,width=0.65\textwidth}
\vspace{.3cm}
\caption{${\rm COP}$ against cooling rate for the 3-level quantum 
absorption cycle, at 3 values of $T_e$. System parameters 
are fixed at:  $T_c=0.1$, $T_h=0.2$,
$\Delta_{31}=1$, $\Lambda_h=\Lambda_c=\Lambda_e=\epsilon=0.001$
and $s_c=1$.
}
\label{fig:cop2}
\end{figure}

The characteristic chiller curve is graphed in Fig \ref{fig:cop2}. 
With $T_e>T_h$, there is no irreversibility that undermines 
efficient slow operation 
($\Delta_{21}\rightarrow \Delta_{21}^{\rm max}$). 
Hence the ordinate intercept of each curve can approach the 
reversible Carnot limit for absorption cycles of \cite{gordon00}
\beq\label{eq:copab}
{\rm COP}_{\rm Carnot}=\frac{\frac{1}{T_h}-\frac{1}{T_e}}
{\frac{1}{T_c}-\frac{1}{T_h}}\,.
%{\rm COP}_{Carnot}=\frac{1/T_h\, - \, 1/T_e}
%{1/T_c\, -\, 1/T_h}\,.
\eeq
% 

%%%%%%%%%%%%%%%%%%%%%%%%%%%%%%%%%%%%%%%%%%%%%%%%%%%%%
%%%%%%%%%%%%%%%%%%%%%%%%%%%%%%%%%%%%%%%%%%%%%%%%%%%%%%

Sections \ref{se:cr} and \ref{se:qac} addressed the extreme cases 
of the power input to the cooling cycle being either 
pure coherent radiation or purely incoherent and thermal, 
respectively. A combination of the two (when $T_e>T_h$) 
can be calculated directly from the results derived above.

%%%%%%%%%%%%%%%%%%%%%%%%%%%%%%%%%%%%%%%%%%%%%%%%%%%%%%%%%%%%

\section{Summary}\label{se:con}

In approaching the absolute zero, any cooling cycle must 
be dominated by quantum dynamics. At the same time, one should 
be able to probe its behavior with fundamental chiller 
thermodynamics. The 3-level quantum model proposed and explored 
here provides a simple case study. It steps beyond earlier quantum 
refrigeration models in accounting fully for spontaneous emission 
(and spontaneous absorption), and including an environmental bath 
that either mimics actual rate-dependent dissipative mechanisms in 
work-driven chillers when $T_e\le T_h$, or establishes the quantum analog 
of an absorption (purely thermally-driven) chiller when $T_e>T_h$.
All the irreversibilities modeled here and in previous studies 
\cite{kosloff00,geva96} are
effectively heat leaks. The quantum analog of internal friction 
stems from the non-commutation of the field and system Hamiltonians. 
In the limit of week fields  ($\epsilon\rightarrow 0$), 
this internal dissipation grows negligible.
Incorporating the quantum version of friction remains a challenge for a
future study.

Subject to reasonable approximations that retain the 
integrity of the system and its reservoirs, expressions can 
be derived for: (a) the conditions under which a cooling effect 
can be generated, (b) chiller efficiency as a function of cooling 
rate, and (c) the temperature dependence of the fastest rate at 
which molecular systems can be cooled at ultra-low temperatures. 
The relation of this maximum cooling rate to the energy mode density 
has been established, and the basic result has been shown to respect 
the fundamental bound that was recently established for maximum 
cooling rate from the Second and Third Laws.

{\bf Acknowledgments}

J.P.P. acknowledges a Golda Meier Fellowship from the
Hebrew University of Jerusalem.
Work supported by the Israel Science Foundation.
The Fritz Haber Research Center is supported by the Minerva
Gesellschaft f\"ur die Forschung, GmbH M\"unchen, Germany.

%%%%%%%%%%%%%%%%%%%%%%%%%%%%%%%%%%%%%%%%%%%%%%%%%%%%%%%%%%%%

\appendix
\section{Quantum Reduced Dynamics}
\label{ap:rd}

To obtain the evolution equation for the 3-level 
system, we invoke standard tools for quantum open systems: 
the Redfield approach \cite{redfield57,laird91} 
and the secular approximation \cite{vankampen97,davies74}. 
Assuming (a) weak coupling of the system with its bath, 
(b) a weak driving field, (c) rapidly decaying bath correlation
functions, and (d) an initial density matrix in the form of
Eq (\ref{eq:initden}), we can express the evolution equation as:
$$
\dot{\hat{\rho}}_s(t)\,
=\,-\,i\,[\,\H_s+\H_{sf}(t)\, ,\,\hat{\rho}_s(t)\,]\,+\nn
$$
\beqa\label{eq:rse}
&&\{-\bar{G}_{\G_h \G_h\da}(\Delta_{31})\,[\P_{31},\P_{13}
\,\hat{\rho}_s(t)]
-\bar{G}_{\G_h\da \G_h}(-\Delta_{31})\,[\P_{13},\P_{31}
\,\hat{\rho}_s(t)]\}+\{{\rm H.C.}\}+\nn\\
&&\{-\bar{G}_{\G_c \G_c\da}(\Delta_{21})\,[\P_{21},\P_{12}
\,\hat{\rho}_s(t)]
-\bar{G}_{\G_c\da \G_c}(-\Delta_{21})\,[\P_{12},\P_{21}
\,\hat{\rho}_s(t)]\}+\{{\rm H.C.}\}+\nn\\
&&\{-\bar{G}_{\G_e \G_e\da}(\Delta_{32})\,[\P_{32},\P_{23}
\,\hat{\rho}_s(t)]
-\bar{G}_{\G_e\da \G_e}(-\Delta_{32})\,[\P_{23},\P_{32}
\,\hat{\rho}_s(t)]\}+\{{\rm H.C.}\}\,\nn\\
&&
\eeqa
where $\{{\rm H.C.}\}$ denotes the Hermitian conjugate 
of the expression in brackets that immediately precedes it. 
Eq (\ref{eq:rse}) is valid independent of the nature of the baths 
as long as the coupling is weak and the bath correlation 
functions decay quickly.

The influence of the baths is included in the coefficients 
$\bar{G}$,
\beq\label{int1}
\bar{G}_{\G_b \G_b'}(\Delta_b)\,=\,
\int_{0}^\infty\,dt\, e^{i t \Delta_b}\,C_{\G_b \G'_b}\,
\eeq
with $\Delta_b=\Delta_{31},\Delta_{21},\Delta_{32}$ for $b=h,c,e$,
respectively, and $C_{\G_b \G'_b}$ being given by (\ref{eq:corr}).
When $\G_b'=\G_b\da$, $\bar{G}$ can be decomposed as 
\beq\label{eq:coe}
\bar{G}_{\G_b \G_b\da}(\Delta_b)\,=\,
\frac{\bar{C}_{\G_b \G_b\da}(\Delta_b)}{2}\,+\,i\,
\bar{S}_{\G_b \G_b\da}(\Delta_b)\,
\eeq
with $\bar{C}$, $\bar{S}$ being real and $\bar{C}\geq 0$
\cite{geva95,laird91,alicki87}.
With Eq (\ref{eq:coe}), we split each term of Eq 
(\ref{eq:rse}) into two parts related to $\bar{C}$ 
and $\bar{S}$. The $\bar{C}$-related term is
\beq\label{eq:cterm}
-\frac{\bar{C}_{\G_b \G_b\da}(\Delta_b)}{2}\{[\P_{b},\P_{b}\da
\hat{\rho}_s(t)]+[\hat{\rho}_s(t) \P_{b},\P_{b}\da]\}
-\frac{\bar{C}_{\G_b\da \G_b}(-\Delta_b)}{2}\{[\P_{b}\da,\P_{b}
\hat{\rho}_s(t)]+[\hat{\rho}_s(t) \P_{b}\da,\P_{b}]\}\,\nn
\eeq
with $\P_b\,=\,\P_{31}$, $\P_{21}$, $\P_{32}$ for $b=h,c,e$
respectively.
It is readily confirmed that the terms of (\ref{eq:cterm}) 
have the standard Lindblad form that insures complete positivity 
of the dynamics.

For simplicity of notation in Sections \ref{se:3lr}-\ref{se:bm} and 
Appendix \ref{ap:coe}, we introduce the notation

\beq
\lambda_b=\lambda_b(\Delta_b,T_b)=
\bar{C}_{\G_b \G_b\da}(\Delta_b)\,,\;\;\;\;\;\;
\bar{\lambda}_b=\bar{\lambda}_b(\Delta_b,T_b)=
\bar{C}_{\G_b\da \G_b}(-\Delta_b)\,
\eeq
with $\lambda_b$ and $\hat{\lambda}_b$ related by
\beq\label{eq:rel1}
\bar{\lambda}_b=e^{-\Delta_b/T_b}\lambda_b\,.
\eeq

The $\bar{S}$-related term is
\beq
i \bar{S}_{\G_b \G_b\da}(\Delta_{b})
[\P_{b} \P_{b}\da,\hat{\rho}_s(t)]+
i \bar{S}_{\G_b\da \G_b}(-\Delta_{b})
[\P_{b}\da \P_{b},\hat{\rho}_s(t)]\,.
\eeq
$\P_{b}\P_{b}\da$ and $\P_{b}\da \P_{b}$
represent small corrections to the system's energy 
levels, which in the weak coupling limit also turn out 
to be negligible.

Hence, from Eq (\ref{eq:rse}), the time evolution equation for any operator 
(in the Heisenberg representation) can be obtained, and the 
germane results are provided in Section \ref{se:3lr}.

%%%%%%%%%%%%%%%%%%%%%%%%%%%%%%%%%%%%%%%%%%%%%%%%%%%%%%%%%%%%

\section{Coefficients at steady state}
\label{ap:coe}

The coefficients $p_{ij}$ are the solution of a $4\times 4$ 
system of linear equations the coefficients of 
which depend on: (a) the system energy structure, 
(b) the coupling with the field $\epsilon$, and 
(c) the coupling with the baths $\lambda_b$ and
$\bar{\lambda}_b$:
\beqa
p_{11}&=&(c_{10}+c_{11}\,\lc+c_{12}\,\lc^2)/D\,\nn\\
p_{22}&=&(c_{20}+c_{21}\,\lc+\bar{c}_{21}\,\lhc
+\bar{c}_{22}\,\lc \lhc)/D\,\nn\\
p_{33}&=&(c_{30}+c_{31}\,\lc+\bar{c}_{31}\,\lhc+c_{32}\,\lc^2
+\bar{c}_{32}\,\lc \lhc)/D\,\nn\\
p_{32}&=&i\,(c_{c0}+c_{c1}\,\lc+\bar{c}_{c1}\,\lhc)/D\,,
\;\;\;\;\;\;p_{23}=p_{32}^*\,
\eeqa
where the denominator $D$ is
\beq
D=c_{d0}+c_{d1}\,\lc+\bar{c}_{d1}\,\lhc+c_{d2}\,\lc^2
+\bar{c}_{d2}\,\lc \lhc\,.
\eeq
and the $c$ coefficients are given by
\beqa
c_{d0}&=&(\lh+\lae+\lhe)\{(\lh+\lhh)\lhe+\lae\lhh\}
+4\epsilon^2(\lh+ 2 \lhh)\,\nn\\
c_{d1}&=&(\lh+\lae+\lhe)(\lh+\lhh+\lae)+(\lh+\lhh)\lhe+\lae\lhh\
+4\epsilon^2\,\nn\\
\bar{c}_{d1}&=&(\lh+\lae+\lhe)^2+8\epsilon^2\,\nn\\
c_{d2}&=&\lh+\lhh+\lae\,\nn\\
\bar{c}_{d2}&=&\lh+\lae+\lhe\,\nn\\
c_{10}&=&(\lh+\lae+\lhe)\lh \lhe + 4\epsilon^2 \lh\,\nn\\
c_{11}&=&(\lh+\lae+\lhe)(\lh+\lae)+\lh \lhe + 4\epsilon^2\,\nn\\
c_{12}&=&(\lh+\lae)\,\nn\\
c_{20}&=&(\lh+\lae+\lhe)\lae\lhh+4\epsilon^2\lhh\,\nn\\
c_{21}&=&\lae\lhh\,\nn\\
\bar{c}_{21}&=&(\lh+\lae+\lhe)(\lh+\lae)+4\epsilon^2\,\nn\\
\bar{c}_{22}&=&(\lh+\lae)\,\nn\\
c_{30}&=&(\lh+\lae+\lhe)\lhe\lhh+4\epsilon^2\lhh\,\nn\\
c_{31}&=&(\lh+\lae+\lhe)\lhh+\lhe\lhh\,\nn\\
\bar{c}_{31}&=&(\lh+\lae+\lhe)\lhe+4\epsilon^2\,\nn\\
c_{32}&=&\lhh\,\nn\\
\bar{c}_{32}&=&\lhe\,\nn\\
c_{c0}&=& 2\epsilon \lhh (\lae-\lhe)\,\nn\\
c_{c1}&=&-2\epsilon \lhh\,\nn\\
\bar{c}_{c1}&=&2\,\epsilon\,(\lh+\lae-\lhe)\,.
\eeqa

The coefficients $A$, $B$ and $C$ of Eq (\ref{eq:phf}) 
can be expressed as
\beqa
A&=&4\epsilon^2 \lh\lc/D\,\nn\\
B&=&4(\lhh+\lhc)(\lae-\lhe)/D\,\nn\\
C&=&\lh\lc\lae(\lh+\lc+\lae+\lhe)/D\,.
\eeqa
%

%%%%%%%%%%%%%%%%%%%%%%%%%%%%%%%%%%%%%%%%%%%%%%%%%%%%%%%%%%%%%%%%%%%%

%%%%%%%%%%%%%%%%%%%%%%%%%%%%%%%%%%%%%%%%%%%%%%%%%%%%%%%%%%%%%%%%%%%%%

%%%%%%%%%%%%%%%%%%%%%%%%%%%%%%%%%%%%%%%%%%%%%%%%%%%%%%%%%%%%%%%%%%%%%

\end{document}